\documentclass[12pt]{article}%%s%%%% Choose Draft version or final
\usepackage[british]{babel}
\def\AnswerYes{y}
\def\pdflatex{y}                  %%%%%% use pdflatex? (if not, latex)
\def\ShowLabelsVersion{n}         %%%%%% Version with defs. of refs, cites or final
\def\ShowChangesVersion{n}        %%%%%% Version with changes highlighted or final
\def\ShowAnnotationsVersion{n}    %%%%%% Version with annotations or final

\def\feynVersion{n}                %%%%%% Choose whether feynman graphs to be
\usepackage[final]{graphicx}            %%% for pictures and figures
\ifx\pdflatex\AnswerYes
   \usepackage[update,prepend]{epstopdf} % automaticlaly converts .eps to .pdf file
   \usepackage{grffile} % allows graphics files to have "." in filename
\fi
\ifx\pdflatex\AnswerYes
    \usepackage{color}
\else
    \usepackage[dvips]{color}
\fi
\ifx\pdflatex\AnswerYes
    \usepackage{thumbpdf}    %
\else
    \usepackage[dvips]{thumbpdf}
\fi
\ifx\pdflatex\AnswerYes
    \usepackage[ 
        final,
        breaklinks=true,
        colorlinks=true,           % coloured links 
        pdfborder={0 0 1},    % coloured box about link
        citecolor={blue},linkcolor={blue},urlcolor={red},
        citebordercolor={1 0 0},linkbordercolor={0 0 1},urlbordercolor={1 0 0},
        pdfpagemode=UseOutlines,% whether Acrobat opens with
        bookmarks=true,bookmarksopenlevel=4% write .pdf table of contents.
        ]{hyperref}         % Hyperlinks for PDF versions
\else
    \usepackage[dvips,              % dvips,ps2pdf,pdftex driver 
        final,
        breaklinks=true,
        colorlinks=true,      % coloured links 
        pdfborder={0 0 1},    % coloured box about link
        citecolor={blue},linkcolor={blue},urlcolor={red},
        citebordercolor={1 0 0},linkbordercolor={0 0 1},urlbordercolor={1 0 0},
        pdfpagemode=UseOutlines,% whether Acrobat opens with
        bookmarks=true,bookmarksopenlevel=4% write .pdf table of contents.
         ]{hyperref}         % Hyperlinks for PDF versions
\fi
\usepackage[numbers,sort&compress]{natbib}%%% collapses [1,2,3] -> [1-3] etc
\usepackage{multirow}                   %%% tabs over multiple rows
\usepackage{amssymb}
\usepackage{amsmath}
\usepackage{xspace}                    %%% correct spacing self-defined macros
\usepackage{dcolumn}                    %%% Align table columns on decimal point
\usepackage{bm}                        %%% bold math

\usepackage{slashed}
\ifx\feynVersion\AnswerYes
   \usepackage{feynmp} % for Feynman diagrams; nice definitions after \begin{document}
\fi
\usepackage{pbox}

\usepackage{epsfig}

\usepackage[textsize=scriptsize]{todonotes}

\ifx\ShowLabelsVersion\AnswerYes
   \usepackage[color]{showkeys} 
   \definecolor{refkey}{gray}{.5}   % slightly gray font
   \definecolor{labelkey}{gray}{.5} % slightly gray font
   
\fi

\ifx\ShowAnnotationsVersion\AnswerYes
   \newcommand{\comment}[1]{{\scriptsize\sffamily\bfseries{#1}}}
   \newcommand{\margin}[1]{\marginpar{\scriptsize\sffamily\bfseries{#1}}}
   \newcommand{\drafty}{\textbf{Draft version \today} \hfill}
\else
   \newcommand{\comment}[1]{}
   \newcommand{\margin}[1]{}
   \newcommand{\drafty}{}
\fi

\ifx\ShowChangesVersion\AnswerYes
   \usepackage{ulem}
   \newcommand{\delete}[1]{\sout{#1}}            % delete #1 (strike-out)
   \renewcommand{\emph}[1]{\textit{#1}}           % ulem overwrites def of \emph as
\else

   \newcommand{\sout}[1]{}
   \newcommand{\xout}[1]{}

   \newcommand{\delete}[1]{}
\fi

\def\lsim{\mathrel{\rlap{\lower4pt\hbox{\hskip1pt$\sim$}}
    \raise1pt\hbox{$<$}}}         %less than or approx. symbol
\def\gsim{\mathrel {\rlap{\lower4pt\hbox{\hskip1pt$\sim$}}
    \raise1pt\hbox{$>$}}}         %greater than or approx. symbol

\newcommand{\MeV}{\ensuremath{\mathrm{MeV}}}

\newcommand{\ChiEFT}{$\chi$EFT\xspace}

\newcommand{\NXLO}[1]{N\ensuremath{{}^{#1}}LO\xspace}

\newcommand{\threeHe}{${}^3$He\xspace}

\def\he3{$^3$He}

\def\3d{3-D}

\newcommand{\mytitle}[1]{\begin{center}\LARGE{\textbf{#1}}\end{center}}
\newcommand{\myauthor}[1]{\textbf{#1}}
\newcommand{\myaddress}[1]{\textit{#1}}
\newcommand{\mypreprint}[1]{\begin{flushright}#1\end{flushright}}

\begin{document}
%%%%%%%%%%%%%%%%%%%%%%%%%%%%%%%%%%%%%%%%%%%%%%%%%%%%%%%%%%%%%%%%%%%%%%%%%%%%%%%
%%%%%%%%%%%%%%%%%%%%%%%%%%%%%%%%%%%%%%%%%%%%%%%%%%%%%%%%%%%%%%%%%%%%%%%%%%%%%%%
% This is a nice title page including abstract ....
%

\begin{titlepage}
  \setcounter{page}{0} \mypreprint{
    %%%%%%%%%%%%%%%%%%%%%%%%%%%%%%%%%%%%%
    \drafty
    %%%%%%%%%%%%%%%%%%%%%%%%%%%%%%%%%%%%%
    3rd April 2018 \\
  }
  
%  \vspace*{0.5cm}
  
  \mytitle{Erratum: Investigating Neutron Polarizabilities through Compton Scattering on ${}^3$He}
  
%  \vspace*{0.5cm}

\begin{center}
  \myauthor{Deepshikha Shukla$^{a}$}\footnote{Email:dshukla@rockford.edu}
   \myauthor{Andreas Nogga$^{b,c}$} \footnote{Email:a.nogga@fz-juelich.de}
  \emph{and}
  \myauthor{Daniel R. Phillips$^{c}$} \footnote{Email:phillid1@ohio.edu}
  
  \vspace*{0.5cm}
   \myaddress{$^a$ Department of Mathematics, Computer Science, and Physics, Rockford University, Rockford, IL 61108, USA}\\[2ex]
   \myaddress{$^b$ Institute for Advanced Simulations 4, Institute f\"ur Kernphysik 3, J\"ulich Center for Hadron Physics, and JARA - High Performance Computing, Forschungszentrum J\"ulich, D-52425 J\"ulich, Germany}\\[2ex]
    \myaddress{$^c$ Department of Physics and Astronomy and Institute of Nuclear
    and Particle Physics, Ohio University, Athens, OH 45701, USA}\\[2ex]
  \vspace*{0.2cm}

\end{center}

%\vspace*{0.5cm}

\begin{abstract}
We provide updated predictions for elastic $\gamma {}^3$He cross sections and asymmetries that correct erroneous results we published in
Phys.\ Rev.\ Lett.\  {\bf 98}, 232303 (2007) and  Nucl.\ Phys.\ A {\bf 819}, 98 (2009).
\end{abstract}
\vskip 1.0cm
\noindent
\begin{tabular}{rl}
  % PACS discontinued as of April 2013 Suggested PACS numbers:&
  % \begin{minipage}[t][\height][t]{10.7cm} 02.30.Rz, 02.30.Uu, 11.80.Jy,
  %   13.75.Cs, 14.20.Dh, 21.30.-x, 25.40.Dn, 27.10.+h\end{minipage}\\[4ex]
  Suggested Keywords: &\begin{minipage}[t]{10.7cm} Effective Field Theory,
    few-nucleon systems, Helium-3, Compton scattering,
     nucleon
    polarisabilities, spin polarisabilities, Chiral Perturbation Theory
                    \end{minipage}
\end{tabular}

\vskip 1.0cm

\end{titlepage}

\setcounter{footnote}{0}

\newpage

%%%%%%%%%%%%%%%%%%%%%%%%%%%%%%%%%%%%%%%%%%%%%%%%%%%%%%%%%%%%%%%%%%%%%%%%%%%%%%%
%%%%%%%%%%%%%%%%%%%%%%%%%%%%%%%%%%%%%%%%%%%%%%%%%%%%%%%%%%%%%%%%%%%%%%%%%%%%%%%
%%%%%%%%%%%%%%%%%%%%%%%%%%%%%%%%%%%%%%%%%%%%%%%%%%%%%%%%%%%%%%%%%%%%%%%%%%%%%%%
% Main Body

%%%%%%%%%%%%%%%%%%%%%%%%%%%%%%%%%%%%%%%%%%%%%%%%%%%%%%%%%%%%%%%%%%%%%%%%%%%%%%%
%%%%%%%%%%%%%%%%%%%%%%%%%%%%%%%%%%%%%%%%%%%%%%%%%%%%%%%%%%%%%%%%%%%%%%%%%%%%%%%
\noindent Here we report errors in the results presented in the publications that described our $O(Q^3)$ calculation of elastic Compton scattering from ${}^3$He in chiral perturbation theory
~\cite{Shukla:2007,Shukla:2009}. The analytic formulas for Feynman diagrams in Ref.~\cite{Shukla:2009} are correct. However, the computer code that calculated observables for Compton scattering from ${}^3$He contained the following errors in the implementation of the two-nucleon Compton operator that appears at $O(Q^3)$ in the chiral expansion: 
\begin{enumerate}
\item It failed to include the isospin dependence of this operator: $(\tau^{(1)} \cdot \tau^{(2)} - \tau^{(1)}_z \tau^{(2)}_z)/2$ (see Eq. (59) of Ref.~\cite{Shukla:2009}). This factor is 1 for ``deuteron"-like pairs with two-body isospin $T=0$ and $-1$ for $T=1$ $np$ pairs. (It is zero for $pp$ and $nn$ pairs.) 

\item There was a factor of two missed when coding the piece of the operator that produces transitions between pairs where the two-nucleon spin, $S$, changes from $S=0$ to $S=1$. 

\item There was a mistake in the implementation of the $S=0$ to $S=1$ piece of diagram E.
\end{enumerate}
None of the errors occur in the earlier code that computes Compton scattering from the deuteron, since there there is only a $T=0$, $S=1$ $np$ pair. 

%%%%%%%%%%%%%%%%%%%%%%%%%%%%%%%%%%%%%%
\begin{figure}[!htbp]
\begin{center}
  \includegraphics[width=0.45\linewidth]
  {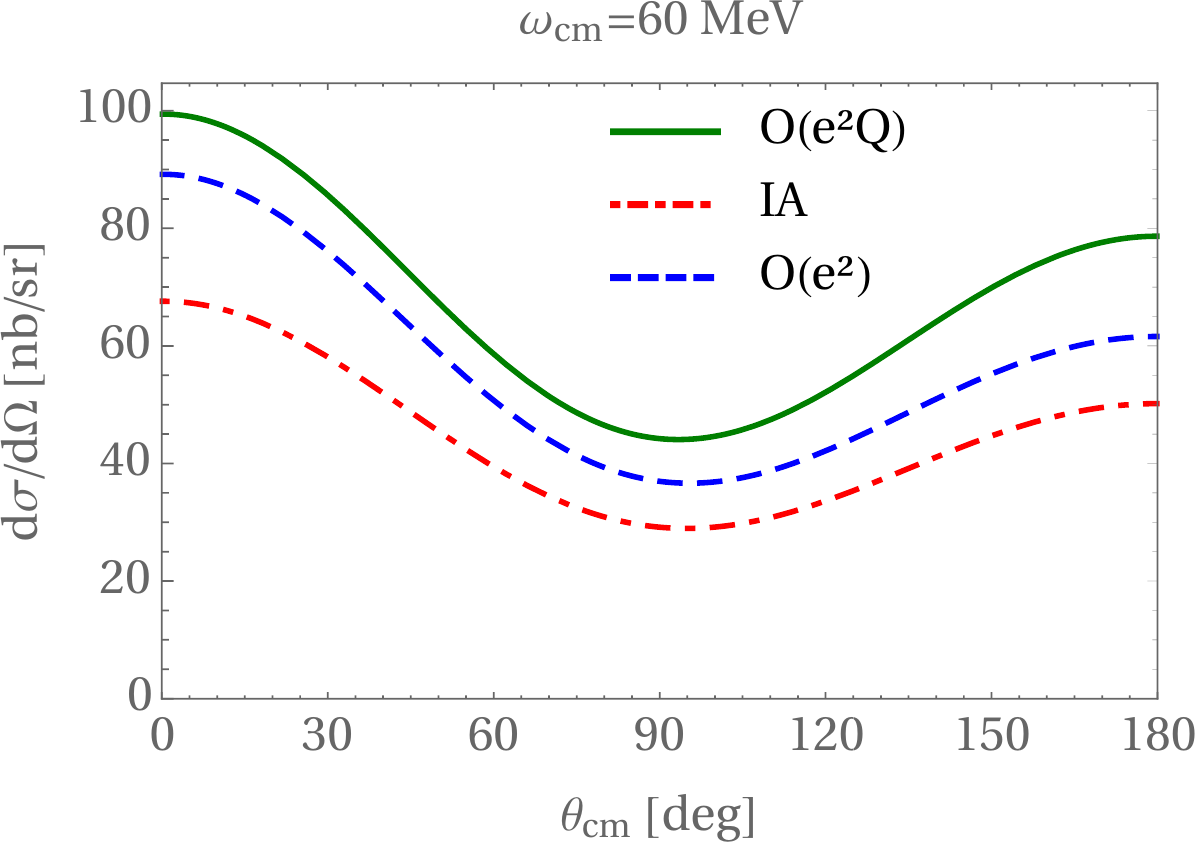}
  \includegraphics[width=0.45\linewidth]
  {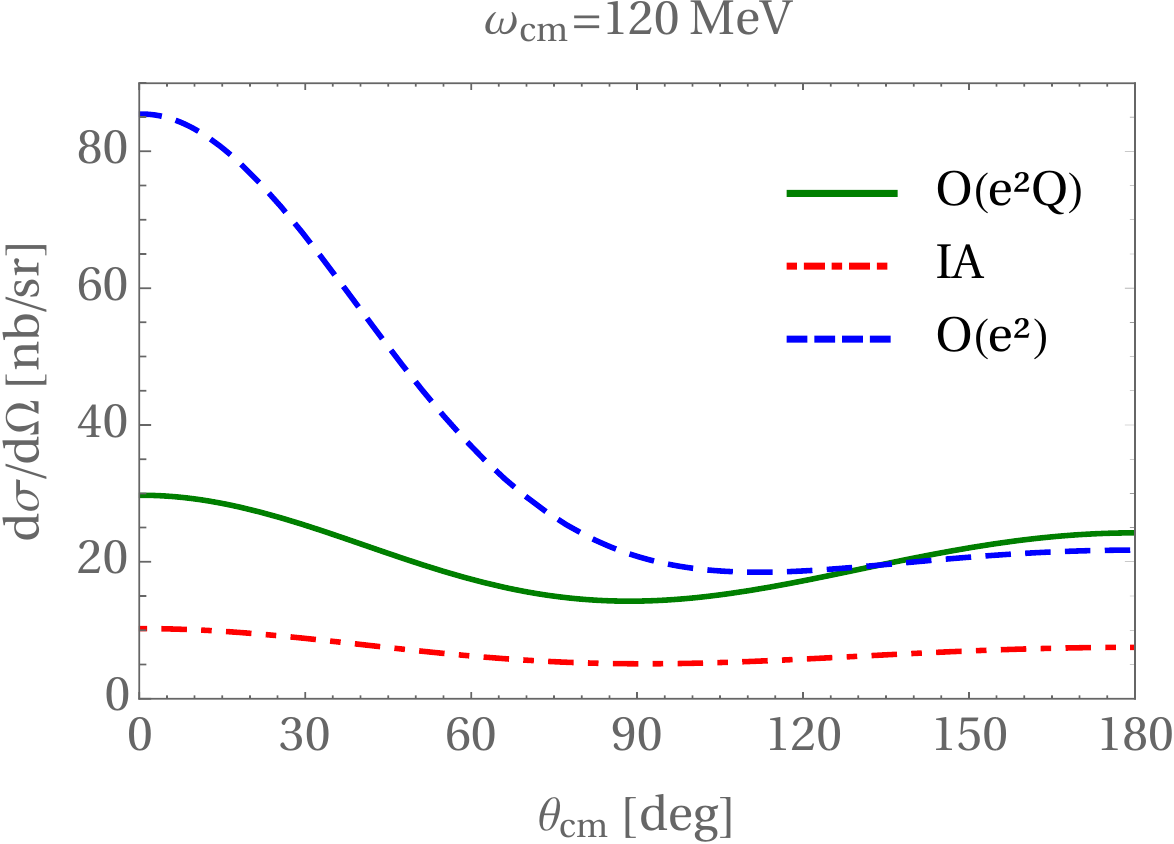}
  \caption{(Color online) The $\gamma$\threeHe cross section at $\omega_{\rm cm}=60$ and $120\;\MeV$. The blue dashed line is the $O(e^2)$ result, the red dash-dotted line represents the impulse-approximation cross section, and the green solid line is the full result at $O(e^2 Q)$.}
\label{fig:dcs}
\end{center}
\end{figure}
%%%%%%%%%%%%%%%%%%%%%%%%%%%%%%%%%%%%%%

The first mistake has the largest consequences for elastic $\gamma {}^3$He scattering. The others only affect parts of the elastic matrix elements that involve small components of the ${}^3$He wave function. But rectifying the isospin factor increases the prediction for the $\gamma ^3$He cross section markedly: see Fig.~\ref{fig:dcs} which replaces Fig.~1 from Ref.~\cite{Shukla:2007} and shows the differential cross section at 60 and 120 MeV for an $O(e^2)$ ($\equiv O(Q^2)$) calculation, an impulse-approximation calculation, and the $O(e^2 Q)$ ($\equiv O(Q^3)$) calculation.  In this, and all the subsequent calculations in this erratum, the 
\ChiEFT Idaho NN potential at \NXLO{3}~\cite{En03} combined with an \NXLO{2} \ChiEFT three-nucleon force~\cite{No06} was used to generate the cross sections. This was also the default choice in Ref.~\cite{Shukla:2007}. The $O(e^2)$ and impulse-approximation results in Fig.~\ref{fig:dcs} are the same as in Refs.~\cite{Shukla:2007,Shukla:2009}. The corrected $O(e^2 Q)$ result is approximately 10\% larger at $\omega_{\rm cm}=60$ MeV than the old, incorrect one. It is 50\% (at forward angles) to 60\% (at backward angles) larger at $\omega_{\rm cm}=120$ MeV.

\begin{figure}[!htbp]
\begin{center}
  \includegraphics[width=0.45\linewidth]
  {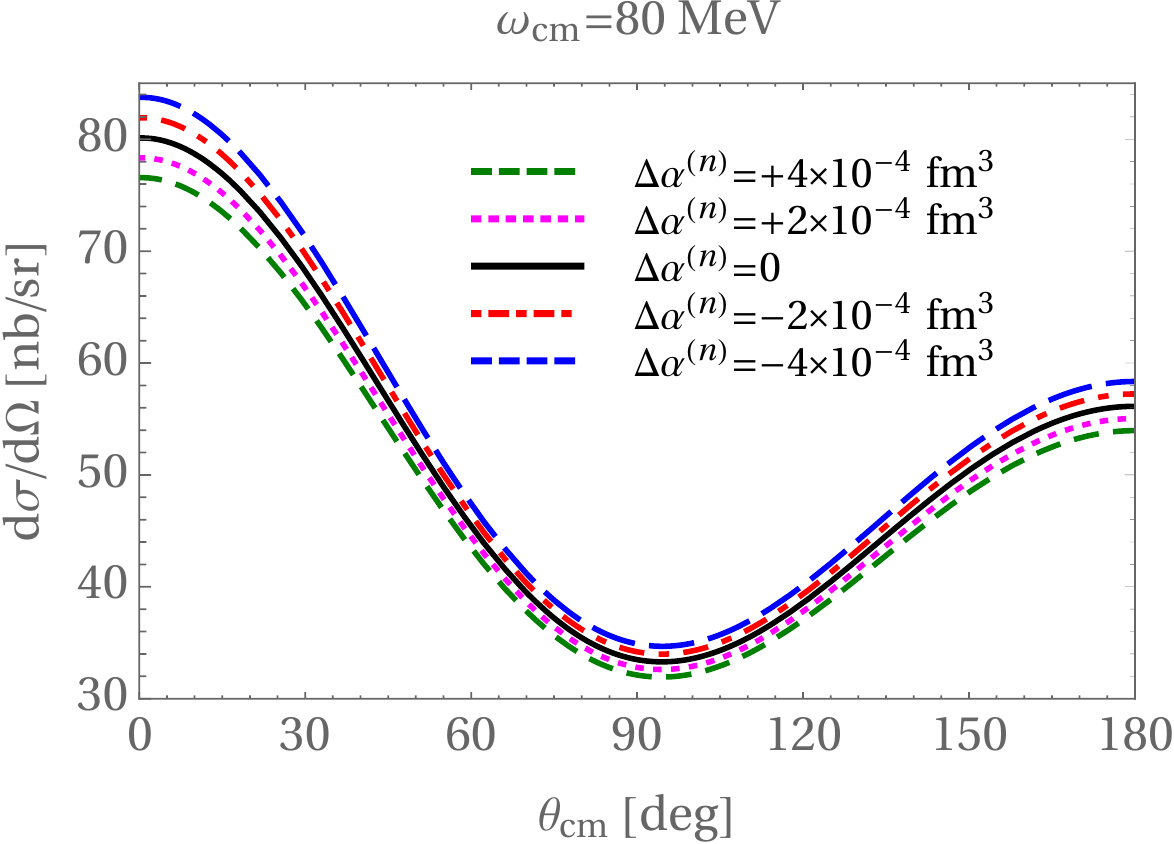}
  \includegraphics[width=0.45\linewidth]
  {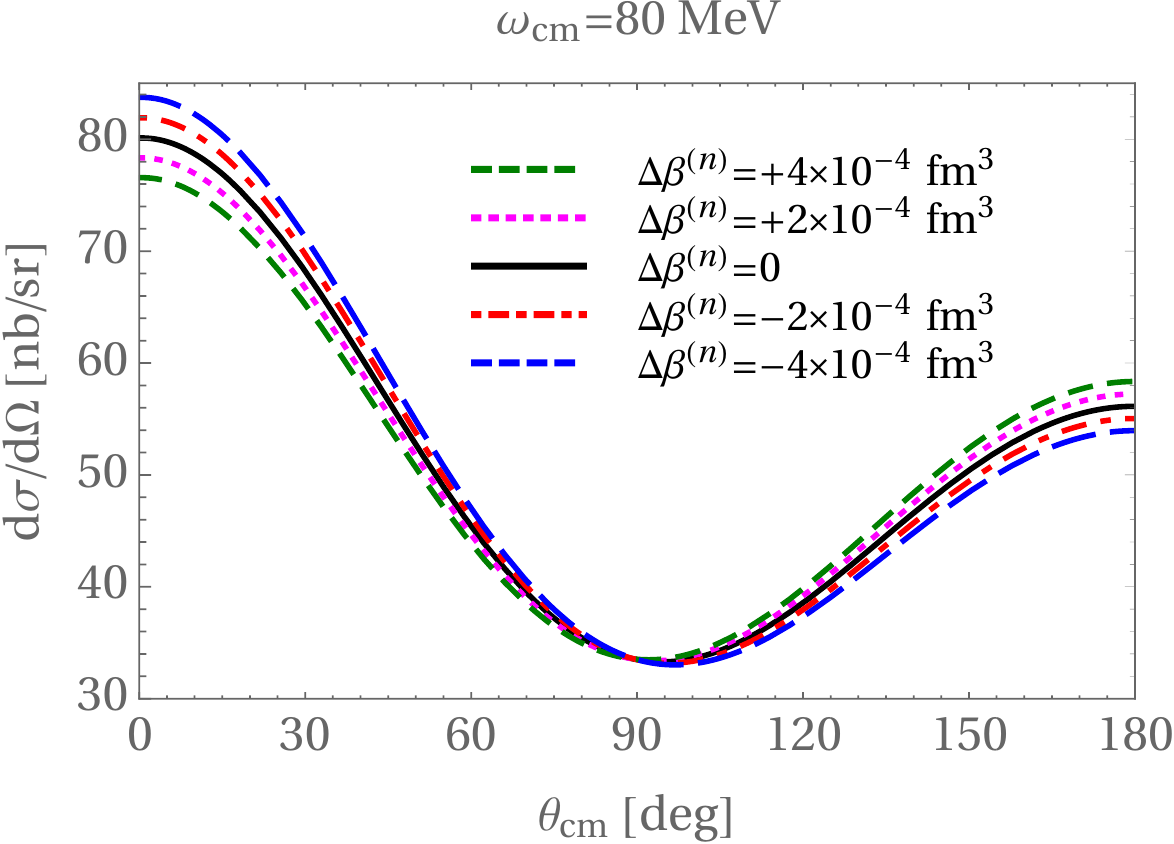}
  \caption{(Color online) The $\gamma$\threeHe cross section at $\omega_{\rm cm}=80$ MeV, for various modifications to the static scalar dipole polarizabilities of the neutron. The left panel shows the sensitivity to changes in $\alpha^{(n)}$, and the right panel the sensitivity to changes in $\beta^{(n)}$. Curves as shown in the panel legends.}
\label{fig:polssensitivity}
\end{center}
\end{figure}

Meanwhile, Fig.~\ref{fig:polssensitivity} shows the corrected results for Fig.~2 of Ref.~\cite{Shukla:2007}, which displayed the sensitivity of the $\gamma {}^3$He differential cross section to changes in the neutron's electric and magnetic static dipole polarizabilities. The pattern and relative size of the effect is very similar to that seen in the original calculation, although the cross section we now obtain is a little larger in absolute size.

%%%%%%%%%%%%%%%%%%%%%%%%%%%%%%%%%%%%%%
%%%%%%%%%%%%%%%%%%%%%%%%%%%%%%%%%%%%%%

%%%%%%%%%%%%%%%%%%%%%%%%%%%%%%%%%%%%%%
\begin{figure}[!htbp]
  \includegraphics[width=0.45\linewidth]
  {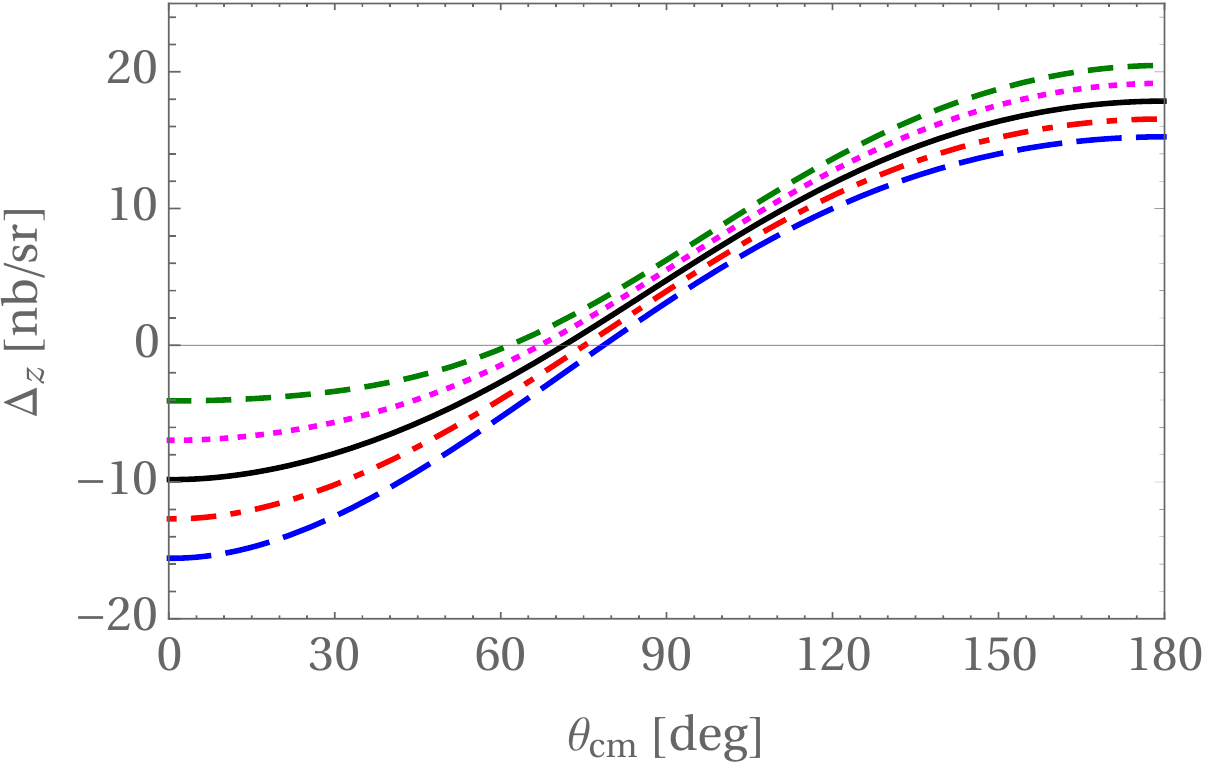}
  \includegraphics[width=0.45\linewidth]
  {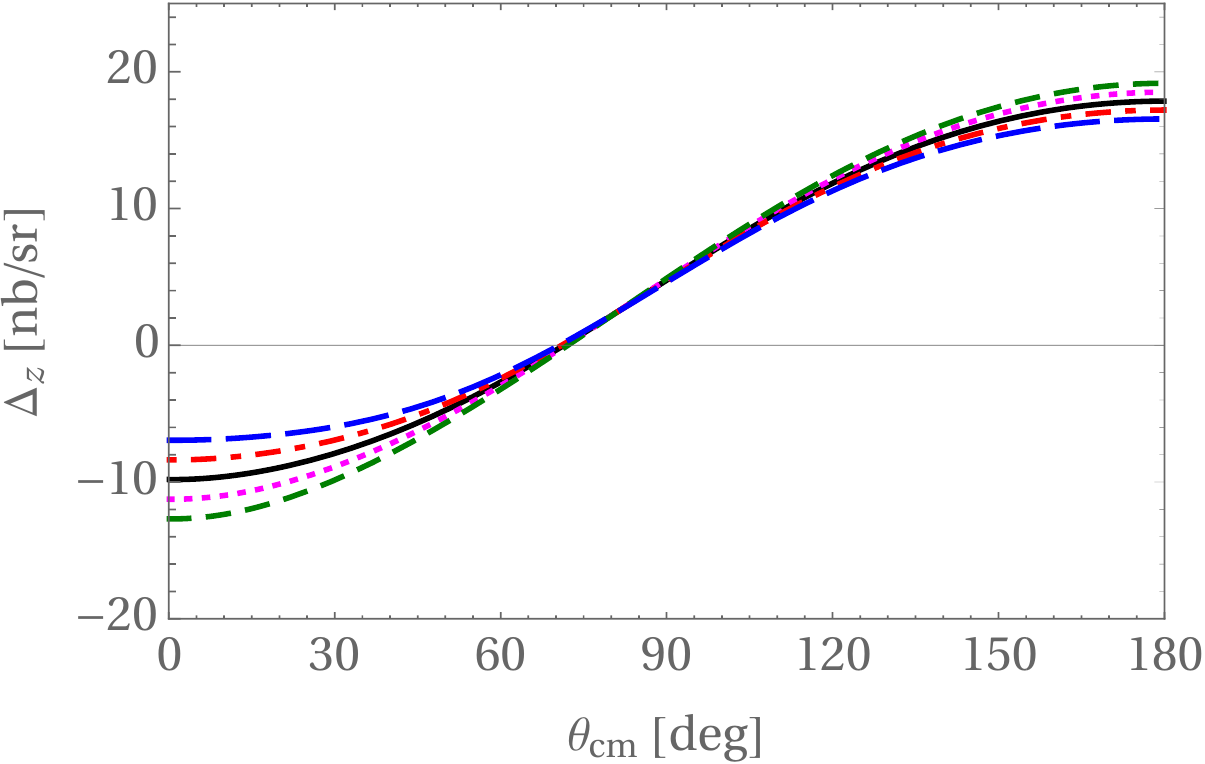}\\
    \includegraphics[width=0.45\linewidth]
  {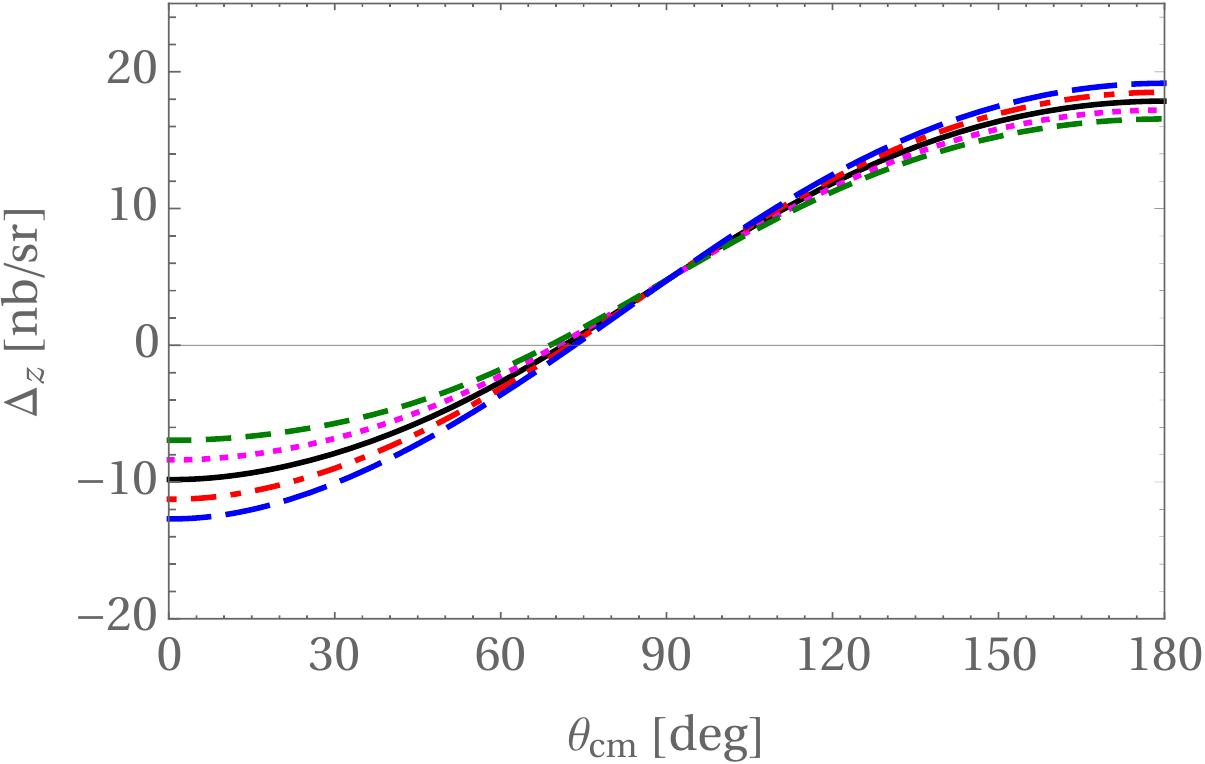}
 \vskip -4cm
 \hskip 9cm
   \includegraphics[width=0.25\linewidth]{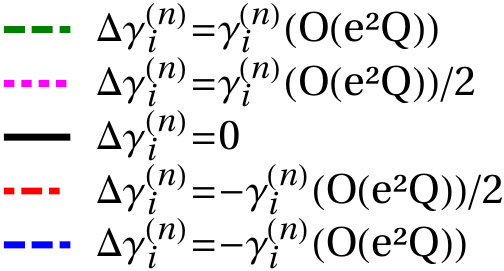}
   \vskip 2cm
  \caption{(Color online) Differences of polarized $\gamma$\threeHe cross sections at $\omega_{\rm cm}=120$ MeV, as obtained under shifts of different static spin dipole polarizabilities of the neutron. The shifts range from $+ 100$\% to -$100$\% of the $O(e^2 Q)$ values. The top left panel shows the sensitivity to changes in $\gamma_1^{(n)}$, the top right panel to $\gamma_2^{(n)}$, and the bottom left panel to $\gamma_4^{(n)}$. Curves as shown in the legend.}
\label{fig:Deltaz}
\end{figure}
%%%%%%%%%%%%%%%%%%%%%%%%%%%%%%%%%%%%%%

%%%%%%%%%%%%%%%%%%%%%%%%%%%%%%%%%%%%%%
\begin{figure}[!htbp]
\begin{center}
  \includegraphics[width=0.45\linewidth]
  {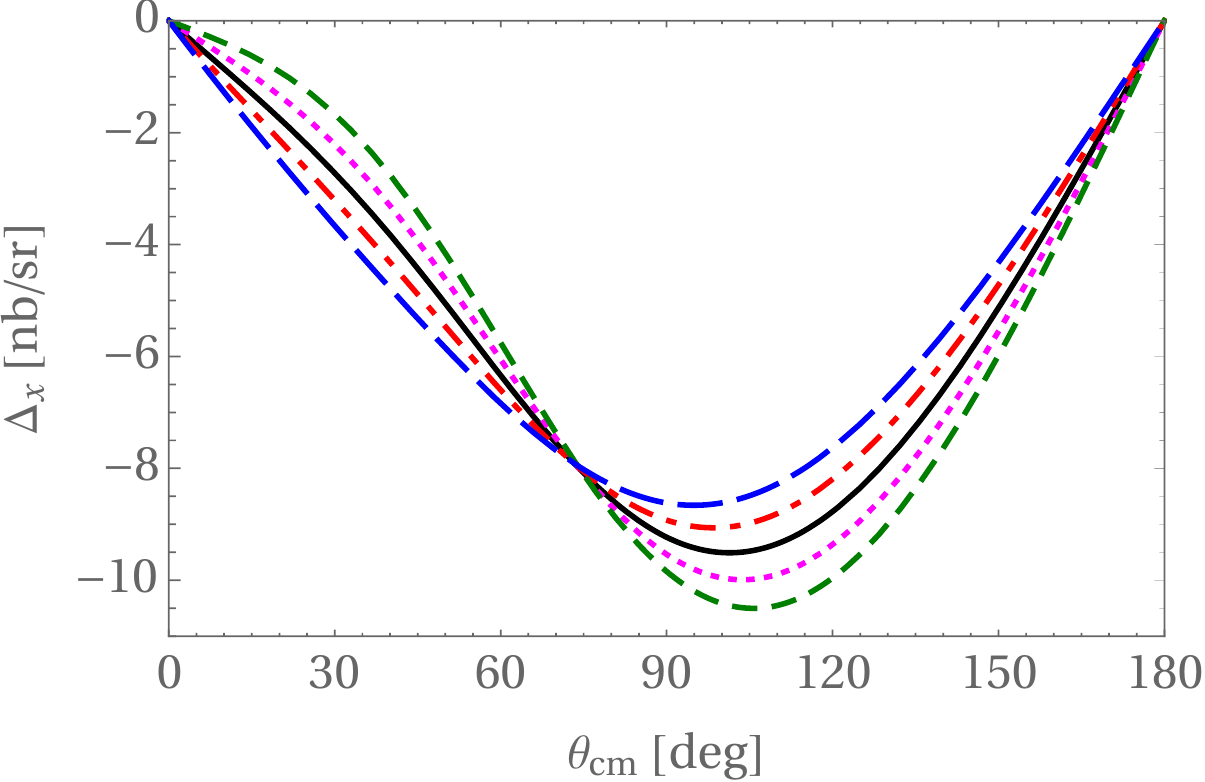}
  \includegraphics[width=0.45\linewidth]
  {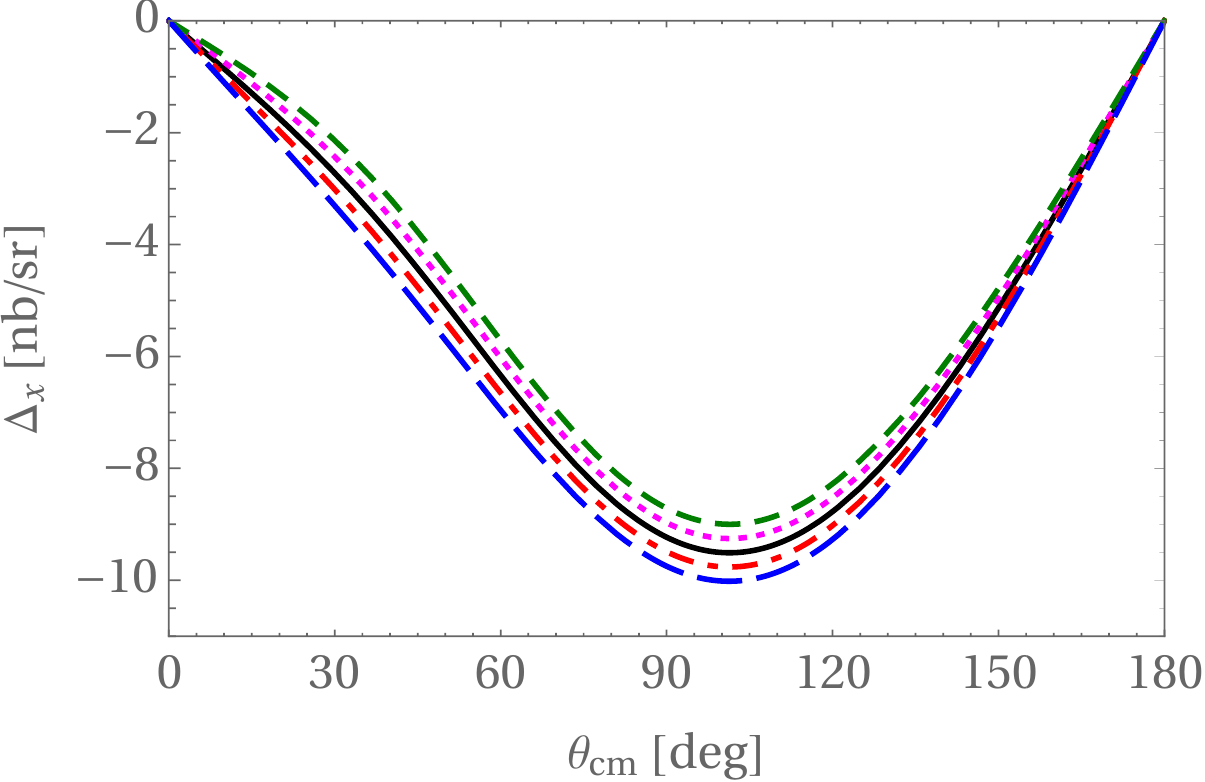}
  \caption{(Color online) Differences of polarized $\gamma$\threeHe cross sections at $\omega_{\rm cm}=120$ MeV, as obtained under changes to the neutron spin polarizability $\gamma_1^{(n)}$ (left panel) and $\gamma_4^{(n)}$ (right panel). Legend as in Fig.~\ref{fig:Deltaz}.}
\label{fig:Deltax}
\end{center}
\end{figure}
%%%%%%%%%%%%%%%%%%%%%%%%%%%%%%%%%%%%%%

Fig.~\ref{fig:Deltaz} then shows corrected results for the difference of differential cross sections for ${}^3$He nuclei polarized parallel and anti-parallel to the beam direction, for a right-circularly polarized photon beam ($\Delta_z$). And Fig.~\ref{fig:Deltax} shows a similar cross-section difference but for target polarization in the scattering plane and perpendicular to the photon beam ($\Delta_x$). In each case the sensitivity to particular neutron spin polarizabilities is displayed.
These two figures for $\Delta_x$ and $\Delta_z$ correct Figs.~3 and 4 from Ref.~\cite{Shukla:2007}. The trend of the effect is very similar to that seen in the old results, but the cross section differences are now $\approx 20$\% larger in Fig.~\ref{fig:Deltaz} and $\approx 40\%$ larger in Fig.~\ref{fig:Deltax}. 
  
None of these changes in our results modify the qualitative conclusions of Refs.~\cite{Shukla:2007,Shukla:2009}. It remains true that:
\begin{enumerate}
\item The Compton cross section on ${}^3$He is significantly larger than that on deuterium. Indeed, this conclusion is strengthened after the corrections discussed here are made.

\item The Compton differential cross section on ${}^3$He has a larger (in absolute terms) sensitivity to the neutron electric and magnetic dipole polarizability than does the deuteron Compton differential cross section.

\item Elastic Compton scattering on a polarized ${}^3$He target produces cross-section differences whose sensitivity to neutron spin polarizabilities is similar to those for a free neutron. 
\end{enumerate}

Lastly, we point out that in defining the asymmetries plotted in Ref.~\cite{Shukla:2009} we divided by the sum of the cross sections for different initial-state polarization of the ${}^3$He nucleus, rather than by the average. Although this was clearly defined in Ref.~\cite{Shukla:2009} (see Eqs. (69) and (71) therein) it is a possible source of confusion when comparing with other works. 

A concurrent publication~\cite{Margaryan:2017} presents corrected versions of all $O(Q^3)$ [$\equiv O(e^2 Q)$] calculations that were presented in  Ref.~\cite{Shukla:2009}. It  also displays the results for the $\gamma {}^3$He observables in an EFT with the $\Delta(1232)$ as an explicit degree of freedom. 

\section*{Acknowledgments}

We thank Harald Grie\ss hammer for generating the graphs shown here, Arman Margayan for identifying the errors and correcting the code, and Bruno Strandberg for his help running the corrected code. This work was supported by the US DOE under grant number DE-FG02-93ER-40756.

\end{document}